# Inter-site pair superconductivity: origins and recent validation experiments

Dragan Mihailovic

*Jozef Stefan Institute, Jozef Stefan International Postgraduate School and CENN Nanocenter, Jamova 39, SI-1000 Slovenia*

## Abstract

The challenge of understanding high-temperature superconductivity has led to a plethora of ideas, but 30 years after its discovery in cuprates, very few have achieved convincing experimental validation. While Hubbard and t-J models were given a lot of attention, a number of recent experiments appear to give decisive support to the model of real-space inter-site pairing[1-3] and percolative superconductivity[4] in cuprates. Systematic measurements of the doping dependence of the superfluid density $\rho_s$ show a linear dependence on $T_c$ - rather than doping - over the entire phase diagram[5], in accordance with the model's predictions. The doping-dependence of the anomalous lattice dynamics of in-plane Cu-O mode vibrations observed by inelastic neutron scattering[6], gives remarkable reciprocal space signature of the inter-site pairing interaction[1] whose doping dependence closely follows the predicted pair density. Symmetry-specific time-domain spectroscopy shows carrier localization, polaron formation, pairing and superconductivity to be distinct processes occurring on distinct timescales throughout the entire superconducting phase diagram. The three diverse experimental results confirm non-trivial predictions made more than a decade ago by the inter-site pairing model in the cuprates, remarkably also confirming some of the fundamental notions mentioned in the seminal paper on the discovery of high-temperature superconductivity in cuprates[7].

## Introduction

The possible relevance of polaron physics was mentioned in the 1986 paper on the discovery of HTS by Bednorz and Müller[7]. Reference was made to a paper by Hock, Nickisch and Thomas[8] who discussed itinerant small Jahn-Teller (JT) polarons. The idea of small superlight (but non-JT) bipolarons was later pursued also by Sasha Alexandrov, where the doping dependence was attributed to a crossover from Bose-Einstein condensation to BCS superconductivity in the metallic overdoped state[9-11]. As an alternative model, paying detailed attention to experimental observations, we proposed the existence of inter-site bipolarons. Some relevant experimental facts in this direction were apparent already by 1990: (i) the presence of doped holes on in-plane oxygens, which suggests dynamical symmetry-breaking around O sites[12], (ii) the anomalous mid-zone phonon dispersion along the $(\xi,0,0)$ direction in the Brillouin zone reported originally by Reichardt and Pintschovius at the APS March meeting in 1990, and later explored in more detail by Egami et al[13], and Reznik et al[14], and (iii) the evidence for broken inversion symmetry from pyroelectricity[15], (iv) a polaronic signature in the mid-infrared[16] accompanied by symmetry-breaking local modes[17] and (v) the possible presence of broken-symmetry pairs above $T_c$ from early $A_{1g}$- symmetry Raman experiments on $YBa_2Cu_3O_{6.9}$ [18]. (This $A_{1g}$ scattering was later confirmed as anomalous and unexplained within standard superconductivity theories [19]). These essential data on carrier localization and symmetry breaking led to the proposal of a new paradigm for the presence of localized polarons co-existing with itinerant fermions[20-22]. Mihailovic and Müller[23] in their review explicitly highlighted the simultaneous presence of (bosonic) bipolarons and itinerant fermions in stripes, as well as the essential role of a long-range Coulomb interaction. Remarkably, the possible coexistence of localized (polarons) and itinerant states in cuprates was discussed on theoretical grounds in the early paper on by Gork'ov and Sokol[24]. Important data in the development of the inter-site model came from ESR[25,26], X-ray absorption fine structure[27] and magnetic susceptibility measurements as often discussed by Alex Müller[28]. A specific proposal of symmetry-based inter-site interaction soon followed thereafter[1,2].

The formal formulation of the inter-site polaron model is based specifically on the observed symmetry-breaking in-plane Cu-O phonon anomalies observed in inelastic neutron scattering experiments which I proposed at the time to be a direct consequence of pair formation [1]. Later Monte-Carlo simulations of the inter-site pairing model[3,29] predicted inter-site pair density which follows $T_c$, rather than being proportional to doping. The model calculation also made the prediction that with increasing doping the metallicity systematically increases with the formation of ever larger itinerant clusters. These predictions have now been apparently confirmed by two decisive experiments which we discuss below.

The importance of percolation appeared in the original "discovery" paper by Bednorz and Müller perhaps in the context of enhanced $T_c$s in granular aluminium, which was of interest at the time. Modifying the percolation idea to one of phase coherence percolation (PCP) between inter-site pairs[4], leads to a simple quantitative prediction of the upper limit of $T_c$ for superconductivity, with the pairing energy and pair size as the only parameters, which we will show below in the context of inter-site pairing.

This research direction was very much outside the box (or "tent") of the mainstream of electronic mechanisms which were probably motivated by the report of the absence of an isotope effect in optimally doped $YBa_2Cu_3O_7$ and $EuBa_2Cu_3O_7$ by Battlogg early in 1987[30]. Later, with extensive and careful isotope effect measurements this result at optimum doping was found to be correct[31], but anomalous. Moreover, it was shown to be consistent with the polaron picture[31].

In this paper, I first summarize the model's main predictions, highlight new and remarkable experimental confirmations of inter-site pairing and then discuss predictions of carrier localization, pair-formation and superconductivity from the viewpoint of recent time-domain experiments.

## The predictions of the JT-C model.

The JT-like interaction responsible for the formation of inter-site pairs was formulated by Kabanov and Mihailovic in 2000[1,32,33], which described the pairing interaction of a single polaron on the $CuO_2$ plane, taking into account that stretching along one direction necessarily means contraction along the other (Fig.1). This is formally expressed in terms of St. Venan's strain compatibility equations [1,3,29,34]. These considerations give an interaction with a Jahn-Teller-like term, in the sense that two (degenerate) configurations are possible in the ground state along x and y axes respectively. Extending the model to multiple polarons, and thus including an essential long range Coulomb repulsion term, the JT-C Hamiltonian is given by[3,29,35]:

$$H_{JT-C} = \sum_{i,j} -V_{JT}(i-j)S_i^z S_j^z + \sum_{i,j} V_C(i-j) Q_i Q_j$$

where $V_{JT}$ is the strain coupling with the lattice and $V_C = \frac{e^2}{\epsilon_0 a m}$ is the long-range Coulomb interaction in which $\epsilon$ is the static electric constant, $e$ is the elementary charge, $a$ is the lattice constant. $S^z = \pm 1, 0$ are the pseudospin operators, and $Q_i = S_i^{z^2}$. The application of compatibility conditions ensure that other surrounding bonds must also deform slightly. Although the Cu-O in-plane bond stretch is the dominant displacement, other bonds also change, as indicated schematically in Fig. 1B). Thus four-fold symmetry breaking is dominant, but the accompanying displacements mean that the local symmetry is much lower, perhaps as low as $C_1$, in which case all excitations associated with pair formation are reduced to the totally symmetric representation. Experimentally this means that A-symmetry excitations are dominant, accompanied with "nematicity" (i.e. loss of four-fold symmetry), while $B_{1g}$ and $B_{2g}$ -like excitations appear as remnants of the underlying tetragonal symmetry. An additional effect is the loss of inversion symmetry, which has been observed experimentally[15].

Quantitative predictions of the polaron ordering were made with a Monte-Carlo simulation of the JT-C Hamiltonian by perfomed by Mertelj et al[3,29] (see also contribution by Kabanov in this volume). The main results of the model are summarized in Fig. 2. The model predicts polaron aggregation into pairs and stripes driven by the interplay of attractive strain and repulsive Coulomb interaction. Remarkably, pairs are seen to be present over a range of doping which coincides with the superconducting region of the cuprate phase diagram. Single polarons appear quite distinct localized states at low doping, and are dominant in the "pseudogap" state, persisting into the non-superconducting part of the phase diagram. Their localization is observed in time-resolved optical experiments. Stripes and larger aggregates appear at higher doping levels and dominate on the overdoped side of the phase diagram, naturally explaining the origin of the two components observed experimentally. The fermions within stripes give rise to band-like behavior, increasing metallic conductivity, the development of a Fermi surface and the appearance of quantum oscillations at low temperature.

It is worth pointing out that in the JT-C model, the lattice strain plays a major role in determining the pair phase diagram shown in Figs. 2. The pairing energy scale $V_{JT}$ depends on the local deformation potential for the Cu-O bond. A systematic measurement of the electron-lattice strain coupling relaxation rate of many superconductors with ultrafast spectroscopy shows a direct linear relation between strain and electron-phonon coupling[36] $\lambda_{ep}$. However, the maximum $T_c$ achievable in different compounds depends non-monotonically on the strain or $\lambda_{ep}$. The empirical fact is that there is an optimum Cu-O bond length which maximizes the pairing susceptibility near $a_{opt.} \simeq 1.92$ Å, showing that $\lambda_{ep}$ alone cannot control $T_c$. It is inherent in the JT-C model that the competition between the local strain caused by the carriers' presence and inter-site Coulomb repulsion between inter-site carriers governs the phase diagram, pair density and $T_c$. Decreasing the Cu-O distance increases $\lambda_{ep}$, but also increases Coulomb repulsion, leading to a non-monotonic dependence of $T_c$ on the Cu-O bond length[37]. This phenomenon has escaped understanding till now.

With real space bosonic singlet inter-site pairs, we proposed that superconductivity arises through phase coherence percolation between them[4]. According to the phase diagram in Fig. 2, on the underdoped side the percolation threshold at $T_c$ corresponds to superconductor-insulator-superconductor (SIS) tunneling, while on the overdoped side, the threshold corresponds to SNS tunneling through metallic regions. In Fig. 3 we show the calculated $T_c$ as a function of doping, based on the Monte-Carlo calculation results such as those shown in Fig. 2. To obtain the data in the plot, I used the Monte-Carlo simulations performed by Mertelj for different temperatures at different doping levels (see Fig. 7 in [29]], and then counted the number of pairs in each point on the phase diagram. Fig. 3 shows the number of pairs (in colour scale) as a function of temperature and doping. The $T_c$ line (indicated in Fig. 3) corresponds to the calculated threshold reached at the volume fraction appropriate for 2D pair percolation[4] showing the ubiquitous dome-like $T_c$ curve as a function of doping.

## New experimental evidence

**The doping dependence of superfluid density**. A very important new systematic experimental result by Bozovic et al.,[5] building on the original Uemura plot[38] substantially changed the perception of theory in the HTS field[39]. A measurement of penetration depth as a function of doping in a very large number of La$_{2-x}$Sr$_x$CuO$_4$ samples establishes that $T_c$ scales linearly with the phase stiffness $\rho_{s0}$ in the limit $T \to 0$ limit, which is proportional to the superfluid density $n_s$ via $\rho_{s0} = \frac{\hbar^2}{4k_B}\left(\frac{n_s}{m^*}\right)$, as shown in Fig. 4. This means that the pair density (which we equate with $n_s$) is proportional to $T_c$ throughout the entire phase diagram. This result is incompatible with the prediction of the majority of models of HTS so far, including BCS, as noted by Zaanen[39]. On the other hand, the observed dependence of $T_c$ on pair density naturally arises within the JT-C model as illustrated in Figs. 2 and 3. Comparing the calculated $T_c$ on the basis of the JT-C model (Fig. 3) as a function of doping with the experimental data on the superfluid density, one obtains remarkable agreement. The JT-C model thus naturally explains why $\rho_s$ decreases on the overdoped side, while the normal state conductivity (metallicity) keeps increasing with increasing doping.

Thus, combining this result of Bozovic et al[5] with the large body of literature discussing the Fermi-liquid like properties from systematics of the transport in the normal state, and convincing data on low-temperature quantum oscillations on both sides of optimum doping[40,41], one arrives at the conclusion that the carriers which exhibit the Fermi liquid behaviour are not the ones that give rise to superconductivity. Instead, as the JT-C model predicts that they are continuously exchanged with the carriers in pairs through fluctuations. The importance of these fluctuations in the critical region at the onset of superconductivity was recently discussed from a time-dynamics point of view by Madan et al.[42].

**The doping dependence of the neutron anomaly.** New data by Reznik and co-workers[6] lend further support to the idea that the neutron anomaly (i.e. the anomalous phonon linewidths along $(\xi,0,0)$ and $(0,\xi,0)$) is related to pairing. They report that the width of the anomalous in-plane "half-breathing mode" follows $T_c$ and shows a similar dependence on doping as $n_{pairs}$ and $\rho_{s0}$, as shown in Figure 5. The position of the anomaly at $\xi = 0.25 \sim 0.3$ reciprocal lattice units (r.l.u.) in the Brillouin zone along the bond axes is proposed to be associated with the ionic displacements resulting from pairing of two holes on opposite sides of a Cu ion (Fig. 1). The JT-C model predicts that the linewidth will follow $n_{pairs}$ and the anomaly will vanish if there are no pairs present. While first principles modelling of the lattice dynamics associated with the inter-site pairing fluctuations, single polarons and clusters (stripes) is computationally challenging, and has yet to be performed, we can make some qualitative statements on the basis of the existing model. The lattice distortion of single polarons will not show this particular anomaly at $\xi \simeq 1/4$ r.l.u., which will disappear at low doping. On the overdoped side, bigger fermionic aggregates (stripes or clusters) will change the q-vector of the Cu-O mode and introduce other modes with other displacements, washing out the signature of pairing. The presence of the Cu-O anomaly at this particular $q-$vector is thus specific to pair formation. The authors[6] discuss in detail other possible explanations and note that other existing theories of HTS do not give any satisfactory explanation for the observed anomaly, and particularly its doping and $q$- dependence.

**Evidence for distinct single polaron localization and pseudogap formation.** A time-domain investigation of the relaxation timescales of the different excitations (polarons, pseudogap excitations, superconducting quasiparticles) shows that they can be distinguished on the basis of symmetry, and temperature dependence in optical pump probe experiments. Recent experiments show rotational symmetry breaking associated with the pseudogap and superconducting gap are both dominantly $A_{1g}$ symmetry, but contain admixtures of $B_{1g}$ and $B_{2g}$ components respectively [43], which help in identifying them by symmetry. The observed symmetry breaking associated with pairing[43,44] is in full agreement with the JT-C model's predictions mentioned earlier. Experiments using multiple pulses which were specially designed to reveal the relaxation timescales of different excitations show a rather unexpected result that single carrier localization (single polarons) in BiSCO takes place in $\tau_{polaron} \sim 250\,fs$, while the PG state recovery time (suggested to be associated with pairing) is nearly 3 times longer, $\tau_{pseudogap} \sim 600\,fs$ [44]. Such distinct behavior can be attributed to single particle localization, pairing and aggregation dynamics respectively and may also be seen in the JT-C model predictions[3,29], as discussed above and summarized in Fig. 2. In contrast, the superconductivity (with an accompanying $B_{2g}$ symmetry signature) is established with a timescale of $\tau_{sc} = 3 \sim 5\,ps$ [45]. The superconductivity appears from fluctuations on the same timescale as the quasiparticle recombination lifetime $\tau_{QP}$, which implies that the recovery is governed by Rothwarf-Taylor bottleneck dynamics[46]. The temperature dependence of the single particle excitations and pairs by time-resolved spectroscopy is also important[42]. Experiments show that the latter, associated with superconducting fluctuations are present at temperatures up to $\sim 1.3\,T_c$.

An important fundamental issue in time-domain is to consider whether the pairs live for long enough for phase coherence to be established. Comparing the measured lifetimes for the pairing

($\tau_{PG} \sim 600\ fs$) with the Josephson time ($\tau_J \sim 300 fs$), which characterizes the phase coherence timescale we see that this condition clearly is fulfilled in Bi$_2$Sr$_2$CaCu$_2$O$_8$ [42].

## Discussion

The JT-C model extends the basic ideas of Mott physics with the addition of a long-range Coulomb repulsion, and explicit consideration of the symmetry breaking effect of localized bipolarons indicated experimentally by various observations. This important difference leads to specific symmetry-breaking phenomena and the simultaneous presence of localized carriers (single polarons and bipolarons) and itinerant fermions. Within this model, the spin degrees of freedom are secondary- the AF order is destroyed by doping very quickly, and only fluctuations remain. The experimental fact that singlet pairing is observed means that the inter-site pairs have $S = 0$, in agreement with ESR experiments [25]. The strain effectively has *d*-wave symmetry, just as the more commonly considered $B_{1g}$ zone-boundary e-p interaction. But there is a crucial difference, the JT-C interaction is centered mid-zone near $q = (1/4,0,0)$ r.l.u., not at the zone boundary.

The two-component paradigm of coexisting itinerant and localized states has recently been applied also to other systems[47], theoretically discussing the coexistence of excited polarons and metastable itinerant states in photoexcited metals. Under conditions when the characteristic phonon mode frequency is smaller than the electron hopping, a semiclassical argument predicts an energy barrier between delocalized and localized states[48]. The 2-component paradigm experimentally expounded and theoretically developed for the cuprates thus appears to be of wider interest. The mechanism may also be relevant in pnictide superconductors, where a similar non-monotonic variation of $T_c$ on bond angle implies an interplay of lattice strain and inter-site Coulomb repulsion leading to an optimum inter-particle distance for pairing. In principle, it should be expected to be relevant for any superconductor whose coherence length is comparable to the unit cell size, the Sulphur hydrides being an obvious case. As superconducting $T_c$ s gradually approach room temperature, and coherence lengths become ever shorter, the physics introduced by the JT-C model may serve as a guide to understanding pairing in such materials.

A final remark is in order, regarding the overwhelming importance of the Coulomb interaction, as emphasized by P.W. Anderson. In some sense pairing and stripe formation within the JT-C model is indeed driven by Coulomb repulsion. Without it, polarons would phase separate, preventing pairing. However, it is clear that the simple Hubbard model with only on-site Coulomb repulsion does not capture the essential physics.

## *Concluding remarks*

Ultimately, a useful superconductivity theory is one which demonstrates predictive power, not only regarding $T_c$, but makes unexpected and non-trivial predictions in advance of experiments. The linear relation between pair density and $T_c$, and anomalous behavior of the inelastic neutron scattering finite wavevector phonon anomaly were such predictions by the JT – Coulomb model and were not generally expected on the basis of more common models at the time, particularly the Hubbard model or the $t-J$ model. While the majority of the popular opinion is that *somehow* the Hubbard model should describe the physics of HTS, these new observations are clearly outside its scope. Specifically, as pointed out by Zaanen[39], the dependence of $T_c$ on superfluid density is outside the Hubbard model's predictions as well as other currently popular models, with the exception of inter-site pairing.

At a very memorable private dinner in Geneva at the occasion of the M2S meeting in 2015, Alex Müller remarked that in his opinion HTS was essentially solved with intersite polarons. To my complaint some time ago that the JT-C model did not have very wide acceptance, he commented that a good theory is like a deposit in the bank: experimental proof would come at some time in the future, sooner or later. The persistent development of the inter-site model against significant opposition by the mainstream HTS community and journal editors would not have been possible without numerous discussions and persistent encouragement by K. Alex Müller.

## *Acknowledgments*

The original work reviewed here was performed in close collaboration with Viktor V. Kabanov, and in the case of MC simulations, with Tomaž Mertelj and Joaquin Miranda-Mena.

## *Figure captions*

Figure 1. The inter-site (bi)polaron on a CuO$_2$ plane. A) Its size is defined by the wavevector of the inelastic neutron scattering anomaly $(\xi, 0,0)$, where $\frac{\pi}{\xi} \simeq 1.2$ nm as shown. The corresponding displacements of the Cu-O mode have different magnitude and phase on different lattice sites, as indicated schematically by arrows. Their intersite pair symmetry must conform to compatibility conditions[29], which means that while the inter-site pair breaks four-fold rotational symmetry of the tetragonal lattice leading to a Jahn-Teller-like symmetry breaking, the lattice must also experience smaller displacements of other surrounding ions as shown exaggerated in B) for the YBa$_2$Cu$_3$O$_7$ structure.

Figure 2. A) Aggregation of localized polarons into bipolarons and stripes from [29] calculated using a Monte-Carlo simulation for two densities (underdoped, n=0.08 and overdoped with n=0.26) at two different temperatures (t=0.64 and t=0.1 in units of $V_{JT}$). The interaction parameters are given in ref. [3]. Each square (■) corresponds to a polaron. The light (□) and dark (■) squares correspond to symmetry breaking along x or y axes respectively. The similarity with scanning tunnelling microscopy real-space images[49] is striking. B) the particle count *x(j)* as a function of temperature *t* and doping level $n$ of single polarons (*j=1*), pairs (*j=2*), three-polaron clusters (*j=3*) etc. from the JT-C model using

the same parameters shown in panel A. The pair density is shown shaded. Note that single polarons are dominant in the underdoped region. Remarkably, the pair density shows a dome-like dependence on doping. (The lines are a guide to the eye.)

Figure 3. Pair density $n_{pairs}$ as a function of temperature and the critical temperature based on phase coherence percolation calculated from the Monte-Carlo calculation of $n_{pairs}$. The dashed line represents the $T_c$ when phase coherence percolation threshold is reached in two dimensions, i.e. the volume fraction of pairs $F_v = 0.5$.

Figure 4. The experimental relation between $T_c$ and superfluid density $\rho_{s0}$ from penetration depth measurements[5] (black and green symbols), and the predicted relation of $T_c$ versus the number of pairs $n_{pairs}$ calculated on the basis of a model of percolation [4] between inter-site bipolarons [29] (open circles) from the data in Fig. 3.

Figure 5. A) The doping dependence of the anomalous Cu-O vibration linewidth[6]. B) The position in the Brillouin zone where the anomaly is observed corresponds to the JT-C model. $(\xi, 0,0)$ and $(0, \xi, 0)$ with $\xi \sim$ ¼ in r.l.u..

1. Mihailovic, D. & Kabanov, V. V. Finite wave vector Jahn-Teller pairing and superconductivity in the cuprates. *Phys Rev B* **63,** 054505– (2001).
2. Kabanov, V. V. & Mihailovic, D. Manifestations of mesoscopic Jahn-Teller real-space pairing and clustering in $YBa_2Cu_3O_{7-delta}$. *Phys Rev B* **65,** 212508 (2002).
3. Mertelj, T., Kabanov, V. & Mihailovic, D. Charged particles on a two-dimensional lattice subject to anisotropic Jahn-Teller interactions. *Phys Rev Lett* **94,** 147003 (2005).
4. Mihailovic, D., Kabanov, V. & Muller, K. The attainable superconducting T-c in a model of phase coherence by percolating. *Europhys Lett* **57,** 254–259 (2002).
5. Bozovic, I., He, X., Wu, J. & Bollinger, A. T. Dependence of the critical temperature in overdoped copper oxides on superfluid density. *Nature* **536,** 309–311 (2016).
6. Park, S. R. *et al.* Evidence for a charge collective mode associated with superconductivity in copper oxides from neutron and x-ray scattering measurements of $La_{2-x}Sr_xCuO_4$. *PRB* **89,** 020506 (2014).
7. Bednorz, J. G. & Müller, K. A. Possible high Tc superconductivity in the Ba–La–Cu–O system. *Z. Physik B - Condensed Matter* **64,** 189–193 (1986).
8. Hock, K. H., Nickisch, H. & Thomas, H. Jahn Teller effect in Itinerant Electron Systems - The Jahn-Teller Polaron. *Helvetica Physica Acta* **56,** 237–243 (1983).
9. Alexandrov, A. S. Bose–Einstein condensation of strongly correlated electrons and phonons in cuprate superconductors. *J Phys-Condens Mat* **19,** 125216 (2007).
10. Alexandrov, A. S. & Mott, N. F. Bipolarons. *Rep Prog Phys* **57,** 1197–1288 (1994).
11. Alexandrov, A. New Theory Of Strong-Coupling Superconductors And High-Temperature Superconductivity Of Metallic Oxides. *Phys Rev B* **38,** 925–927 (1988).
12. Nücker, N., Fink, J., Fuggle, J., Durham, P. & Temmerman, W. Evidence for holes on oxygen sites in the high-Tc superconductors $La_{2-x}Sr_xCuO_4$ and $YBa_2Cu_3O_{7-y}$. *Phys. Rev. B* **37,** 5158–5163 (1988).
13. McQueeney, R. J. *et al.* Anomalous Dispersion of LO Phonons in $La_{1.85}Sr_{0.15}CuO_4$ at Low Temperatures. *PRL* **82,** 628–631 (1999).
14. Reznik, D. *et al.* Electron–phonon coupling reflecting dynamic charge inhomogeneity in copper oxide superconductors. *Nature* **440,** 1170–1173 (2006).
15. Mihailovic, D. & Heeger, A. Pyroelectric and piezoelectric effects in single crystal of YBa 2 Cu 3 O 7-δ. *Solid State Communications* (1990).
16. Mihailovic, D. *et al.* Application of the polaron-transport theory to sigma ( omega ) in $Tl_2Ba_2Ca_{1-x}Gd_xCu_2O_8$, $YBa_2Cu_3O_{7-d}$, and $La_{2-x}Sr_xCuO_4$. *Phys. Rev. B* **42,** 7989–7993 (1990).
17. Mihailovic, D., Foster, C., VOSS, K. & Mertelj, T. Anomalous shifts of oxygen-mode frequencies in $La_{2-x}Sr_xCuO_4$, $YBa_2Cu_3O_{7-\delta}$ and $Tl_2Ba_2Ca_{1-x}Gd_xCu_2O_8$ studied by photoinduced infrared absorption and Raman spectroscopy. *Phys Rev B* **44,** 237–241 (1991).
18. Mihailovic, D., Zgonik, M., Copic, M. & Hrovat, M. Phys. Rev. B 36, 3997 (1987): Quasiparticle excitations in the superconducting state observed in light scattering. *Phys Rev B* (1987).
19. Devereaux, T. P. & Hackl, R. Inelastic light scattering from correlated electrons. *Rev Mod Phys* **79,** 175 (2007).
20. Mihailovic, D., Stevens, C., Podobnik, B. & Demsar, J. Evidence for Two-component superconductivity in the femtosecond optical and transient photoconducting …. *Physica C* **282-287,** 186–189 (1997).
21. Stevens, C., Smith, D., Chen, C., Ryan, J. & Podobnik, B. Evidence for Two-Component High-Temperature Superconductivity in the Femtosecond Optical Response of …. *Phys Rev Lett* **78,** 2212 (1997).
22. Mertelj, T., Demsar, J., Podobnik, B., Poberaj, I. & Mihailovic, D. Photoexcited carrier relaxation in YBaCuO by picosecond resonant Raman spectroscopy. *Phys Rev B* **55,** 6061–6069 (1997).
23. Mihailovic, D. & Müller, K. A. in *High-Tc Superconductivity 1996: Ten Years after the Discovery* 243–256 (Springer Netherlands, 1997). doi:10.1007/978-94-011-5554-0_10
24. Gorkov, L. & Sokol, A. Phase Stratification of an Electron Liquid in the New Superconductors. *JETP Lett* **46,** 420–423 (1987).
25. Müller, K. A. The Impact of ESR (EPR) on the Understanding of the Cuprates and their

Superconductivity. *EPR newsletter* **22,** 5–6 (2012).
26. Kochelaev, B., Sichelschmidt, J., Elschner, B. & Lemor, W. Intrinsic EPR in $La_{2-x}Sr_xCuO_4$: Manifestation of Three-Spin Polarons. *Phys Rev Lett* (1997).
27. Bianconi, A., Saini, N. L., Lanzara, A. & Missori, M. Determination of the Local Lattice Distortions in the $CuO_2$ Plane of $La_{1.85}Sr_{0.15}CuO_4$. *Phys Rev Lett* **76,** 3412–3415 (1996).
28. Müller, K. A., Zhao, G.-M., Conder, K. & Keller, H. The ratio of small polarons to free carriers in $La_{2-x}Sr_xCuO_4$ derived from susceptibility measurements. *Journal Of Physics-Condensed Matter* **10,** L291–L296 (1998).
29. Mertelj, T., Kabanov, V. V., Mena, J. M. & Mihailovic, D. Self-organization of charged particles on a two-dimensional lattice subject to anisotropic Jahn-Teller-type interaction and three-dimensional Coulomb repulsion. *Phys Rev B* **76,** 9 (2007).
30. Batlogg, B. *et al.* Isotope Effect in the High- TcSuperconductors $Ba_2YCu_3O_7$ and $Ba_2EuCu_3O_7$. *PRL* **58,** 2333–2336 (1987).
31. Bussmann-Holder, A. & Keller, H. Unconventional isotope effects, multi-component superconductivity and polaron formation in high temperature cuprate superconductors. *Journal of Physics: Conference Series* **108,** 012019 (2008).
32. Boris, A. V. *et al.* Josephson plasma resonance and phonon anomalies in trilayer $Bi_2Sr_2Ca_2Cu_3O_{10}$. *PRL* **89,** 277001 (2002).
33. Kabanov, V. V. & Mihailovic, D. Finite-Wave-Vector Phonon Coupling to Degenerate Electronic States in $La_{2-x}Sr_xCuO_4$. *J Supercond* **13,** 959–962 (2000).
34. Lookman, T., Shenoy, S. R., Rasmussen, K. Ø., Saxena, A. & Bishop, A. R. Ferroelastic dynamics and strain compatibility. *Phys Rev B* **67,** 024114 (2003).
35. Kabanov, V. V., Mertelj, T. & Mihailovic, D. Mesoscopic phase separation in the model with competing Jahn-Teller and Coulomb interaction. *J Supercond Nov Magn* **19,** 67–71 (2006).
36. Gadermaier, C. *et al.* Strain-induced enhancement of the electron energy relaxation in strongly correlated superconductors. *Physical Review X* **4,** 011056 (2014).
37. Rao, C. N. R. & Ganguli, A. K. Structure–property relationship in superconducting cuprates. *Chem Soc Rev* **24,** 1–7 (1995).
38. Uemura, Y. *et al.* Universal correlations between Tc and ns/m (carrier density over effective mass) in high-Tc cuprate superconductors. *Phys Rev Lett* **62,** 2317–2320 (1989).
39. Zaanen, J. Condensed-matter physics: Superconducting electrons go missing. *Nature* **536,** 282–283 (2016).
40. Nakamae, S. *et al.* Electronic ground state of heavily overdoped nonsuperconducting $La_{2-x}Sr_xCuO_4$. *Phys Rev B* **68,** 4 (2003).
41. Badoux, S. *et al.* Universal quantum oscillations in the underdoped cuprate superconductors. *Nat Phys* **9,** 761–764 (2013).
42. Madan, I. *et al.* Separating pairing from quantum phase coherence dynamics above the superconducting transition by femtosecond spectroscopy. *Sci Rep* **4,** 5656 (2014).
43. Toda, Y. *et al.* Rotational symmetry breaking in Bi2212 probed by polarized femtosecond spectroscopy. *Phys Rev B* **90,** 094513 (2014).
44. Madan, I. *et al.* Evidence for carrier localization in the pseudogap state of cuprate superconductors from coherent quench experiments. *Nat Comms* **6,** 6958 (2015).
45. Madan, I. *et al.* Real-time measurement of the emergence of superconducting order in a high-temperature superconductor. *Phys Rev B* **93,** 224520–8 (2016).
46. Kabanov, V. V. & Mihailovic, D. Kinetics of a superconductor excited with a femtosecond optical pulse. *PRL* **95,** 147002 (2005).
47. Sayyad, S. & Eckstein, M. Coexistence of excited polarons and metastable delocalized states in photoinduced metals. *PRB* **91,** 104301 (2015).
48. Kabanov, V. V. & Mashtakov, O. Y. Electron localization with and without barrier formation. *Phys. Rev. B* **47,** 6060–6064 (1993).
49. Hoffman, J. E. *et al.* Imaging quasiparticle interference in $Bi_2Sr_2CaCu_2O_{8+d}$. *Science* **297,** 1148–1151 (2002).

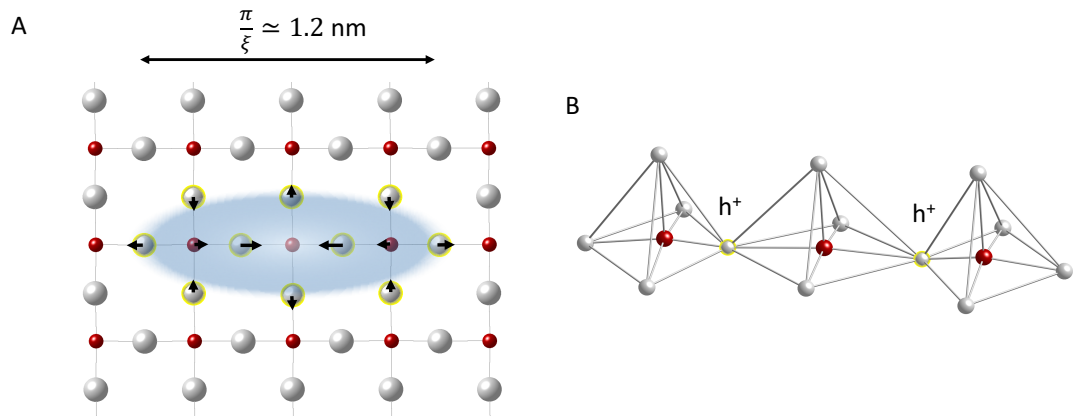

Fig. 1

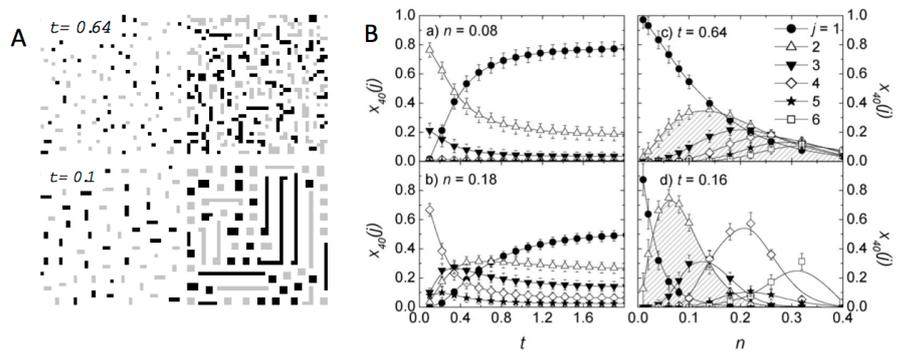

Fig. 2

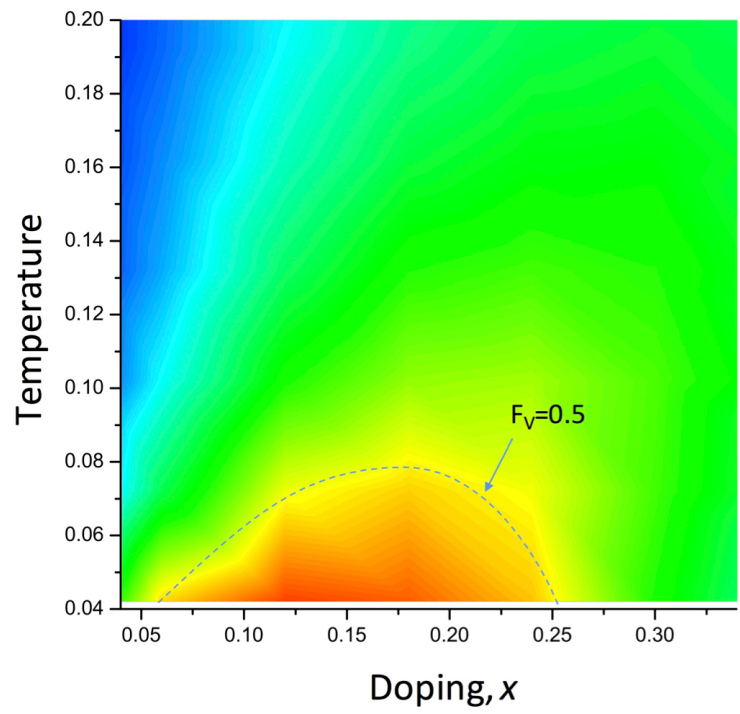

Fig. 3

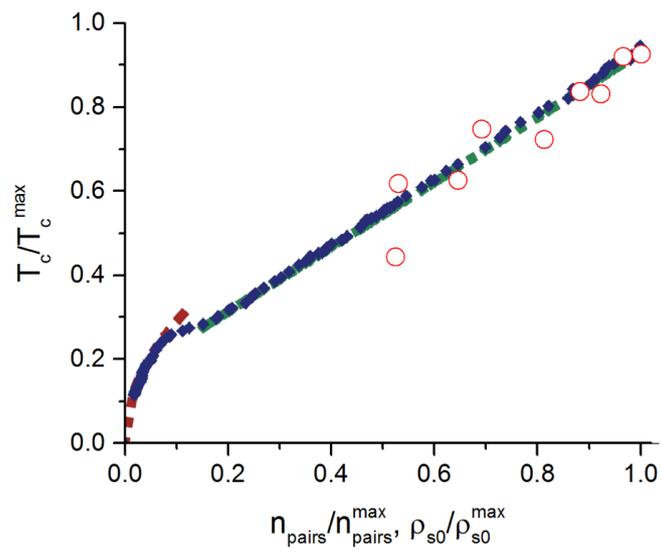

Fig. 4

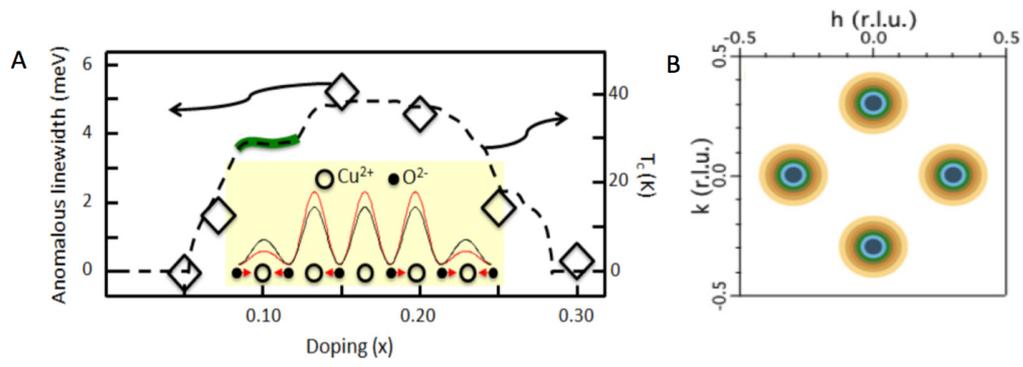

Fig. 5